\newcommand{\simgt}{\,\rlap{\lower 3.5 pt \hbox{$\mathchar \sim$}} \raise
1pt \hbox {$>$}\,}
\newcommand{\simlt}{\,\rlap{\lower 3.5 pt \hbox{$\mathchar \sim$}} \raise
1pt \hbox {$<$}\,}

\def\be{\begin{equation}}
\def\ee{\end{equation}}
\def\ba{\begin{eqnarray}}
\def\ea{\end{eqnarray}}

\def\dm{\delta m^2}
\def\dmeV{\delta m^2_{\rm eV}}

\def\l{\left}
\def\r{\right}

\def\MeV{{\,{\rm MeV}}}
\def\eV{{\,{\rm eV}}}

\documentclass[12pt]{article}
\usepackage[dvips]{epsfig}
\begin{document}

\setlength{\baselineskip}{0.30in}

\newcommand{\nc}{\newcommand}
\newcommand{\num}{\nu_\mu}
\newcommand{\nue}{\nu_{\rm e}}
\newcommand{\nut}{\nu_\tau}
\newcommand{\nus}{\nu_{\rm s}}
\newcommand{\mnus}{m_{\nu_{\rm s}}}
\newcommand{\taus}{\tau_{\nu_{\rm s}}}
\newcommand{\nnt}{n_{\nu_\tau}}
\newcommand{\rnt}{\rho_{\nu_\tau}}
\newcommand{\mnt}{m_{\nu_\tau}}
\newcommand{\tnt}{\tau_{\nu_\tau}}
\newcommand{\bi}{\bibitem}
\newcommand{\rar}{\rightarrow}
\newcommand{\lar}{\leftarrow}
\newcommand{\lrar}{\leftrightarrow}
\newcommand{\so}{\, \mbox{\sin}\Omega}
\newcommand{\co}{\, \mbox{\cos}\Omega}
\newcommand{\sotil}{\, \mbox{\sin}\tilde\Omega}
\newcommand{\cotil}{\, \mbox{\cos}\tilde\Omega}
\makeatletter
\def\alt{\mathrel{\mathpalette\vereq<}}
\def\vereq#1#2{\lower3pt\vbox{\baselineskip1.5pt \lineskip1.5pt
\ialign{$\m@th#1\hfill##\hfil$\crcr#2\crcr\sim\crcr}}}
\def\agt{\mathrel{\mathpalette\vereq>}}

\newcommand{\eq}{{\rm eq}}
\newcommand{\tot}{{\rm tot}}
\newcommand{\M}{{\rm M}}
\newcommand{\coll}{{\rm coll}}
\newcommand{\ann}{{\rm ann}}
\makeatother

\renewcommand{\thefootnote}{\fnsymbol{footnote}}

\ \vspace*{-3.cm}
\begin{flushright}
  {MPI--PhT/2000-37}\\
  {\ }
\end{flushright}

\vspace*{0.3cm}

\begin{center}
\vglue .06in
{\Large \bf {\boldmath Maximum lepton asymmetry from active -- sterile neutrino
oscillations in
the Early Universe.}}
\bigskip
\\{\bf R.~Buras \footnote{e-mail: {\tt rburas@mppmu.mpg.de}}} \\
{\it{Max-Planck-Institut f\"ur Physik (Werner-Heisenberg-Institut)\\
F\"ohringer Ring 6, 80805 M\"unchen, Germany
}}
\\{\bf D.V.~Semikoz \footnote{e-mail: {\tt semikoz@mppmu.mpg.de}}} \\
{\it{Max-Planck-Institut f\"ur Physik (Werner-Heisenberg-Institut)\\
F\"ohringer Ring 6, 80805 M\"unchen, Germany\\
and\\
Institute of Nuclear Research of the Russian Academy of Sciences\\
60th October Anniversary Prospect 7a, Moscow 117312, Russia}}
\\[.40in]
\end{center}

\begin{abstract}
A large lepton asymmetry could be generated in the Early Universe by
oscillations of active to sterile neutrinos with a small mixing angle
$\sin 2 \theta <10^{-2}$. The final order of magnitude of the lepton
asymmetry $\eta$ is mainly determined by its growth in the last stage
of evolution when the MSW resonance dominates the kinetic equations.
In this paper we present a simple way of calculating the maximum
possible lepton asymmetry which can be created. Our results are in
good agreement to previous calculations. Furthermore, we find that the
growth of asymmetry does not obey any particular power law. We find
that the maximum possible asymmetry at the freeze-out of the $n$/$p$
ratio at $T\sim 1 \MeV$ strongly depends on the mass-squared
difference $\dm$: the asymmetry is negligible for $|\dm| \ll 1~ {\rm
eV}^2$ and reaches asymptotically large values for $|\dm| \simgt 50~
{\rm eV}^2$.
\end{abstract}

\newpage
\renewcommand{\thefootnote}{\arabic{footnote}}
\setcounter{footnote}{0}
\section{Introduction}

Neutrino oscillations between muon and tau neutrinos, which were found
recently in the Super-Kamiokande experiment \cite{SK}, prove the
existence of neutrino masses.  This finding does not spoil the Standard
Big Bang Nucleosynthesis (BBN) picture because  experimentally interesting
neutrino masses for all three flavors $\nu_i$, $i = {\rm e}, \mu, \tau$, are
small $m_i \le 1$ eV, in comparison to typical BBN temperatures $T \sim 1$
MeV.  As for oscillations between active flavors, these do not play any
role in BBN because all active  flavors are equally populated.

However, if sterile neutrinos exist, oscillations between active and
sterile states could have significant consequences for BBN. One
possibility is that the total energy density of the Universe is
changed due to excited sterile neutrinos; together with a recent
analysis of observational data \cite{tytler0001318}, which claims that
the effective number of neutrinos can not exceed $3.2$, this effect
can be used to exclude some region of the $(\delta m^2,\sin 2 \theta)$
parameter plane e.g.~for heavy sterile neutrinos \cite{dhrs}.

Another possible influence of sterile neutrinos on BBN arises from the
fact that $\nu_{\rm e}$-$\nu_{\rm s}$ oscillations can create a large
asymmetry between $\nu_{\rm e}$ and $\bar \nu_{\rm e}$ which would
have a direct impact on the $n$-$p$ reactions and thus change the
light element abundances. This case is interesting for small negative
$\delta m^2$ and small $\sin 2 \theta< 10^{-2}$, and the important
result will be the value of the electron neutrino asymmetry at the
freeze-out of the $n$/$p$ ratio at a temperature around 1 MeV.
Actually, the magnitude of $\nu_{\rm e}$ asymmetry at $T=1 \MeV$ is just
an indicator for possible consequences on BBN. For accurate
calculations one would need to use exact non-equilibrium distribution
functions of $\nu_{\rm e}$ and $\bar \nu_{\rm e}$ in the
neutron-proton reactions.  A general discussion of this subject can be
found in Section 2.1 of \cite{KARMEN00}, and details in case of large
asymmetry in \cite{foot00}.

In early works with simplified calculations it was found that the
asymmetry could not reach large final values
\cite{enkvist90,barbieri91}.  However, later it was found that the
asymmetry can increase significantly and reach values
$\mathcal{O}(0.1)$ \cite{foot96}.  This finding was confirmed by
elaborate numerical calculations \cite{fv96,footAP99,bfoot00}.  A main
feature of these results was that the asymmetry grows according to the
power law $T^{-4}$. A contradictory result that the power law was
$T^{-1}$ was suggested in \cite{dolgov99}, but was found to be
erroneous in \cite{we2000,comment00}.

In the present paper we investigate the evolution of the lepton
asymmetry under the assumption that all active neutrinos passing the
resonance are converted to sterile neutrinos, thus forming a
Fermi-Dirac distribution in the sterile sector. This assumption is
correct in the case of pure Mikheyev-Smirnov-Wolfenstein (MSW)
transition \cite{msw}, i.e.~when one can neglect the
collision-integral terms. Although the assumption of full conversion
from active to sterile states may be too strong for certain models,
our result does present the maximum possible asymmetry. The advantage
of our approach is that we do not solve any differential
equations. Thus, our approach is extremely simple and safe.
Furthermore, we consider the effects of various simplifications and
compare our results with previous works, in particular with
\cite{fv96}. Finally, we investigate how the maximum amplitude of the
asymmetry at the temperature 1 MeV depends on initial conditions and
$\dm$, and compare it with a more detailed calculation, \cite{ASF99b}.

\section{Maximal asymmetry at given temperature}

We are considering neutrino oscillations in the Early Universe at
temperatures around $1 \MeV \le T \le 100 \MeV$. In this epoch, the
Universe is a homogeneous and isotropic plasma, and the expansion is
described by the Hubble parameter,
\be
H =\sqrt{\frac{8\pi\rho_{\rm tot}}{3
m_{\rm pl}^2}}~,
\ee
where $m_{\rm pl}=1.22 \times 10^{22} \MeV$ is the Planck mass. The
total energy density is
\be \rho_{\rm tot} = \frac{\pi^2}{30} g_\ast T^4~, \ee
where $g_\ast=10.75$ is the number of effective degrees of freedom during
the epoch of interest. We also suppose that  $\dot T/T = -H $,
which is a good approximation for the models we are considering here.

The neutrino oscillations are described by the density matrix
formalism, see for instance \cite{dolgov81}. However, we will mainly be
interested in the resonance condition, which for $\nu_{\rm a} \leftrightarrow 
\nu_{\rm s}$ oscillations is given by \cite{notzold88}
\be
\frac{\dm \cos 2\theta}{2 E}= - V_{\rm eff}^{\rm a} =-\l( \mp C_1 \eta
G_F T^3 + C_2^{\rm a} \frac{G_F^2 T^4 E}{\alpha}\r),
\label{rescond}
\ee
where $a={\rm e},\mu,\tau$ specifies the type of active neutrino,
$\dm=m_{\rm s}^2 - m_{\rm a}^2$ is the mass squared difference,
$\theta$ is the mixing angle, and $V_{\rm eff}^{\rm a}$ is the
effective potential. The  ``$\mp$'' signs refer to neutrinos and
anti-neutrinos, respectively.  Furthermore, $G_F=1.166\times 10^{-5}$
GeV$^{-2}$ is the Fermi coupling constant, $\alpha=1/137$ is the fine
structure constant, $C_1=0.345$, $C_2^{\rm e}=0.61$ for $\nu_{\rm
e}$-$\nu_{\rm s}$ mixing, $C_2^{\mu, \tau}=0.17$ for
$\nu_{\mu,\tau}$-$\nu_{\rm s}$ and $T$ is the temperature of the
plasma. Sometimes we will use the notation $\dmeV = \dm / {\rm eV}^2$.
Finally, the effective asymmetry is $\eta=2 \eta_{\nu_{\rm
a}}+\eta_0$, where $\eta_{\nu_{\rm a}}$ is the neutrino asymmetry of the 
active
species a, and $\eta_0$ is some small fermion asymmetry from all other
fermion species. For our analysis it is only interesting that $\eta_0$ is of
order of the baryon asymmetry, $10^{-10}$.

Because the temperature is much larger then the neutrino masses,
not only neutrino momentum and energy are
equal, $p=E$, but also that a certain neutrino mode evolves with
constant $y\equiv p/T$. We will therefore use this dimensionless
variable instead of momentum and energy, so $p=E=y T$.

We briefly discuss the evolution of the neutrino densities for small
mixing angles, $\sin2\theta < 10^{-2}$, looking at $\nu_{\rm e}$ for
definiteness. Let us first sketch the reason why the lepton asymmetry
can change at all. Due to the mixing of $\nu_{\rm e}$ with $\nu_{\rm
s}$, the number densities of the $\nu_{\rm e}$ and $\bar\nu_{\rm e}$
can alter. The sterile neutrinos that are thus created have no effect
on $\eta$. Now if $\eta\neq0$ initially, the resonance condition for
the $\nu_{\rm e}$ and $\bar \nu_{\rm e}$ will be different. Therefore
their number densities will change differently, thereby changing
$\eta$. Thus neutrino mixing works as a back-reaction on $\eta$.
The actual change in asymmetry and density takes place close to the
resonance momentum, which is defined by Eq.~(\ref{rescond}),
\be
y_{\rm res}= \sqrt{\l(k_1 \frac{\eta}{T^2}\r)^2+k_2\frac{1}{T^6}} \pm k_1 \frac{\eta}{T^2}~, 
\label{yres}
\ee
where 
\be
k_1=\frac{C_1}{2 C_2^{\rm e}G_F/\alpha}~~~ \mbox{and}~~~ k_2=\frac{|\dm|\cos 2\theta}{2C_2^{\rm e}G_F^2/\alpha}~.
\ee

There are three different stages of lepton asymmetry evolution.
During the first stage the asymmetry is so small that the $k_1$ terms
can be neglected in Eq.~(\ref{yres}). The neutrinos and the
anti-neutrinos simultaneously pass through the resonance. Furthermore,
any initial asymmetry decreases due to the negative back-reaction.

The second stage is the stage of exponential growth of the lepton
asymmetry, which occurs when the back-reaction becomes positive. Here,
the resonance conditions for $\nu$ and $\bar\nu$ are rapidly driven
apart. For definiteness, we will assume that the created asymmetry is
positive, $\eta>0$. Then $y_{\rm res}$ for the $\nu$ increases to very
high momenta where the Fermi distribution is negligible, the resonance
momentum for $\bar\nu$ decreases to small momenta $y_{\rm res} <
1$. At the same time, the asymmetry grows many orders of magnitude and
reaches values of order of $10^{-6}$.

In the third stage the lepton asymmetry grows at a slower rate. There
has been a lot of controversy on this subject, and it has often be
proposed that the lepton asymmetry would grow according to an approximate
power law, $\eta\propto T^{-\alpha}$, where $\alpha=1$-$4$.

We will now show that any power law is only an approximation to the
true asymmetry growth. Our considerations start at the end of the
exponential regime, which happens at temperature $T_i$. The important
variables are the asymmetry $\eta_i(T_i)$ and the resonance momentum
$y_i(T_i,\eta_i)$ at this time. For definiteness, we will assume
$\eta_i>0$, so that we can safely neglect $\nu$ oscillations (with
 $\eta_i<0$ we could neglect $\bar\nu$).

The resonance momentum is governed by two effects: it decreases for
increasing $\eta$ and increases for decreasing $T$. At the end of the
exponential regime, these two effects compensate each other, 
so that at first the
resonance momentum changes very slowly. This plus the fact that
collisional damping is small may let us assume that the momentum modes
passing the resonance in the regime of slow asymmetry growth will
fully excite the sterile sector. For negligible collisional damping,
this is exactly the MSW effect \cite{msw}. The full excitation of the
sterile sector  at resonance has also been stated
in \cite{fv96}.

We can now calculate the asymmetry for a given temperature $T<T_i$:
\be
\eta(T)=\eta_i + \frac{1}{2\zeta(3)}\int\limits_{y_i}^{y_{\rm res}(T)}dy
y^2 f_{\rm eq}(y,\tilde\mu)~.
\label{resint}
\ee
Here, $f_{\rm eq}$ is the equilibrium Fermi distribution function,
given by
\be
f_{\rm eq}(y,\tilde\mu)= \frac{1}{ e^{y-\tilde\mu} +1}~,
\label{feq}
\ee
where $\tilde\mu = \mu/T$ is the dimensionless chemical potential for
the active neutrinos. For $\eta\ll 1$ and large temperatures $T \gg 1\MeV$, 
the chemical potential is
approximately $\tilde\mu\approx - 1.5 \eta$. Therefore, we can safely
neglect it. We will discuss its influence for large $\eta$ later on.
We should mention that $\eta_i$ has its origin in an already weakly
excited sterile sector at momenta $0<y<2$. However, the correction to
Eq.~(\ref{resint}) from this excitation is negligible.

We can now insert Eq.~(\ref{yres}) into Eq.~(\ref{resint}) and solve the
equation for $\eta(T)$. The resulting equation only depends on the
input parameters $\eta_i$ and $y_i$, and on the neutrino parameter
$\dm$. Interestingly, the equation is independent of the mixing angle
$\sin 2\theta$, a fact which stems from the full transition
assumption. The result has another striking feature: even if the
sterile sector is not fully excited during resonance, which happens if
the resonance momentum changes too fast and/or the collisional damping
is important, our result still gives the maximally possible asymmetry
at temperature $T$.

 For small momenta, the Fermi-Dirac
distribution can be approximated by $f_{\rm eq}=1/2$. Then
Eq.~(\ref{resint}) simplifies to
\be
\eta(T)-\eta_i = \frac{1}{12\zeta(3)}\l(y_{\rm res}^3(T)-y_i^3\r)~.
\ee
Furthermore, for large asymmetries and small temperatures, the $k_2$ term in
Eq.~(\ref{yres}) is much smaller than the $k_1$ term, so that
\be
y_{\rm res}\simeq \frac{k_2}{2k_1\eta T^4}~.
\label{late_res}
\ee
\label{yresapp}
Then
\be
\eta^4 + \eta^3\l(\frac{1}{12\zeta(3)} y_i^3- \eta_i\r) = \frac{1}{12\zeta(3)} \l(\frac{k_2}{2k_1 T^4}\r)^3
~.
\label{eta4}
\ee
This equation can be easily  solved in two limits: when $\eta \ll
(\frac{1}{12\zeta(3)} y_i^3- \eta_i)$ we can neglect the first term
on the LHS so that we obtain a $\eta \propto T^{-4}$ power
law. However, when $\eta \gg (\frac{1}{12\zeta(3)} y_i^3- \eta_i)$  the solution 
is $\eta \propto T^{-3}$. Thus the power law will
slowly change in time.

This is also what we expect from the following considerations: the asymmetry
is created close to the resonance momentum. However, when the sterile
sector is fully excited at this momentum, the asymmetry can no longer
be created. Naturally, when $\eta$ does not change, the resonance
momentum will increase due to the decreasing temperature. Thus we
expect a slowly increasing resonance momentum. From
Eq.~(\ref{late_res}) we can then deduce that $\eta$ increases more
slowly than a $\eta \propto T^{-4}$ power law.

We can easily derive an equation similar to Eq.~(14) of \cite{fv96}
by differentiating Eq.~(\ref{resint}) in the temperature $T$:
\be
\frac{d \eta}{dT}= \frac{1}{2\zeta(3)}
{y^2_{\rm res}(T)} f_{\rm eq}(y_{\rm res},0) \frac{dy_{\rm res}}{dT}~.
\label{detadT}
\ee
The only difference between our Eq.~(\ref{detadT}) and Eq.~(14) of
\cite{fv96} is in the usually negligible term proportional to
$f_{\nu_s}$. Furthermore, Eq.~(18) of \cite{fv96}, which can be
derived from their Eq.~(14) directly, is in our notation
\be
\frac{d\eta}{dT}=\frac{-4 \frac{y^3}{2\zeta(3)} f_{\rm eq}}{T+\frac{y^3}{2\zeta(3)} f_{\rm eq}T/\eta}~,
\ee
where we have used the simplification Eq.~(\ref{late_res}).

\begin{figure}
\unitlength1mm
\begin{picture}(121,80)
  \put(20,0){\psfig{file=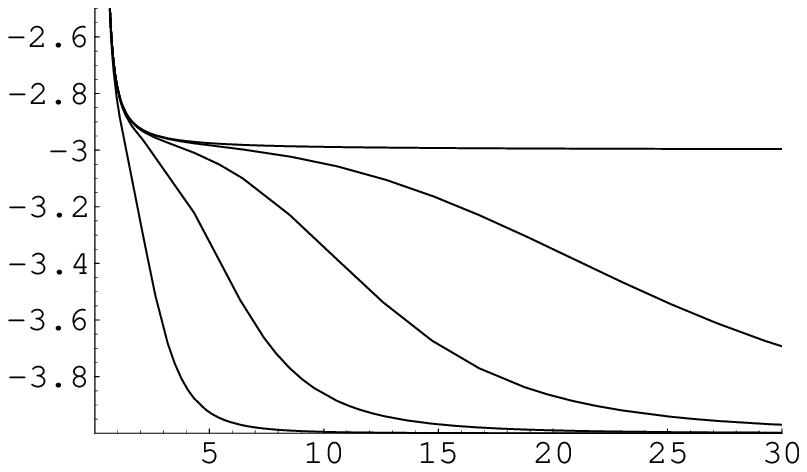,width=10.4cm}} 
  \put(27,66){$\alpha$}
  \put(130,4){$T/(\dmeV)^{1/4}$}
  \put(95,46){$y_i=0$}
  \put(96,30){$0.05$}
  \put(77,20){$0.1$}
  \put(55,20){$0.2$}
  \put(36,15){$0.5$}
\end{picture}
  \caption{Power-law index $\alpha$  as a function of temperature for
  various values of the initial resonance momentum $y_i$. We have used the
  simplified condition for the resonance momentum and have neglected
  the initial asymmetry $\eta_i$. Note that e,g.~$\eta_i\sim
  6 \times 10^{-6}$ would substantially change the line for $y_i\le 0.1$.}
\label{alpha_1}
\begin{picture}(121,80)
  \put(20,0){\psfig{file=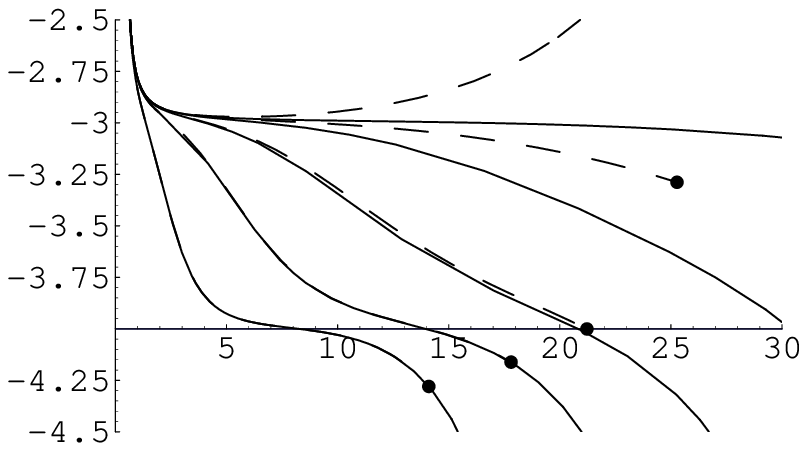,width=10.4cm}} 
  \put(27,66){$\alpha$}
  \put(130,15){$T$}
  \put(95,50){$y_i=0$}
  \put(96,35){$0.05$}
  \put(79,28){$0.1$}
  \put(60,26){$0.2$}
  \put(38,24){$0.5$}
\end{picture}
  \caption{Power-law index  $\alpha$  as a  function of temperature for
  various values of the initial resonance momentum $y_i$. Here we have
  used the exact resonance momentum condition. The solid lines
  represent the case $\eta_i=0$, whereas the dashed lines show the
  effect from non-zero $\eta_i$, in this case $\eta_i=6\times
  10^{-6}$. The dots show starting-points, i.e. $T_i$ and the
  corresponding power $\alpha_i$ (for $\eta_i=0$, $T_i=\infty$).}
\label{alpha_2}
\end{figure}

For illustration, we have plotted the evolution of the ``power''
$\alpha$ which can be derived by the formula
\be
\alpha= \frac{T}{\eta} \frac{d\eta}{dT}
\ee
in Fig.~\ref{alpha_1}. Here, we have already included the effect from
the exact Fermi-Dirac distribution, which becomes important at low
temperatures. We have plotted the evolution for $\dm=1~ \eV^2$ and for
different values of $y_i$. Note that typical values are
$y_i=0.1$-$0.2$. Furthermore, we have set $\eta_i=0$. Typical values
would be $\eta_i=10^{-6}$-$10^{-5}$, but as we can see from
Eq.~(\ref{eta4}), such values are negligible in comparison with the
$y_i^3$ term. Anyhow, we can simulate the effect of $\eta_i$ by
changing $y_i$, see Eq.~(\ref{eta4}). Interestingly, the effect from
$\dm$ can be compensated exactly by rescaling the temperature:
\be
T_{\dm} = (\dm)^{1/4}~ T_{\dm=1 \eV^2}.
\ee

To be still more precise, we need to use the exact Eq.~(\ref{yres})
instead of Eq.~(\ref{late_res}). We have plotted the results in
Fig.~\ref{alpha_2} in analogy to Fig.~\ref{alpha_1}. We see that both
the approximate and the exact resonance condition give the same
results at low temperatures, while at high temperatures the results
are very different; the exact result shows a larger power-law
behavior. This means that there is no real power-law solution in the
MSW dominated region, at least under the assumption that transition is
complete. We should mention that for $\nu_{\mu,\tau}$-$\nu_{\rm s}$
oscillations, the correction due to the exact resonance condition is
somewhat weaker, since $C_2$ is smaller by a factor 4.

We should note that using Eq.~(\ref{late_res}) instead of
Eq.~(\ref{yres}) has no influence on the later evolution of the
asymmetry, i.e. Eq.~(\ref{late_res}) is a good
approximation. This is because the apparently larger increase of
$\eta$ for given $T_i$, $\eta_i$ is compensated by a smaller $y_i$
when using Eq.~(\ref{yres}).
 
In Fig.~\ref{alpha_2} we also show the effect of a non-zero initial
asymmetry. This effect is important only for small initial values of
the resonance momentum $y_{\rm res} < 0.1$.

\section{Asymmetry at the freeze-out of the \boldmath $n/p$ ratio.}

\begin{figure}
\unitlength1mm
\begin{picture}(121,80)
  \put(20,0){\psfig{file=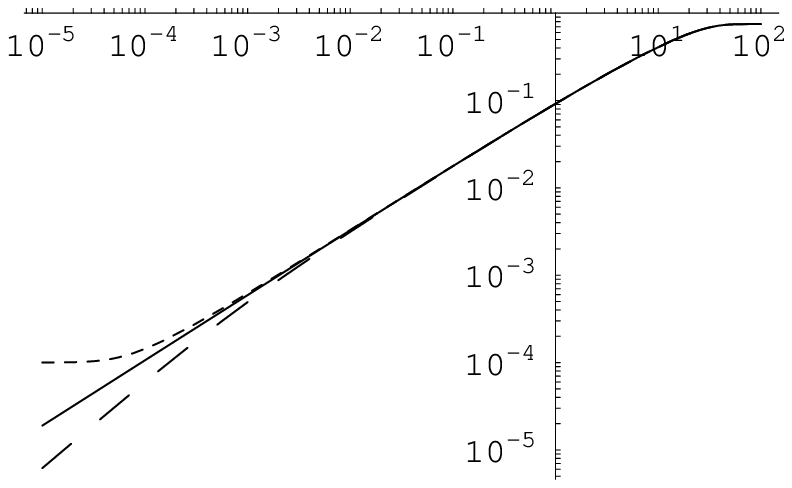,width=10.4cm}} 
  \put(81,0){$\eta_{\rm max}$}
  \put(128,58){$|\dmeV|$}
\end{picture}
  \caption{The maximal possible asymmetry $\eta$ at the freeze-out
temperature of the $n$-$p$ interactions, $T=1\MeV$, as a function of
$|\dmeV|$, for $\eta_i=y_i=0$ (solid line). The initial asymmetry
$\eta_i$ increases the final asymmetry for small $|\dmeV|$ (short
dashed line for $\eta_i=10^{-4}$); the momentum $y_i$ decreases
the asymmetry for small $|\dmeV|$ (long dashed line is for
$y_i=0.2$).}
\label{fnal_1}
\begin{picture}(121,80)
  \put(20,0){\psfig{file=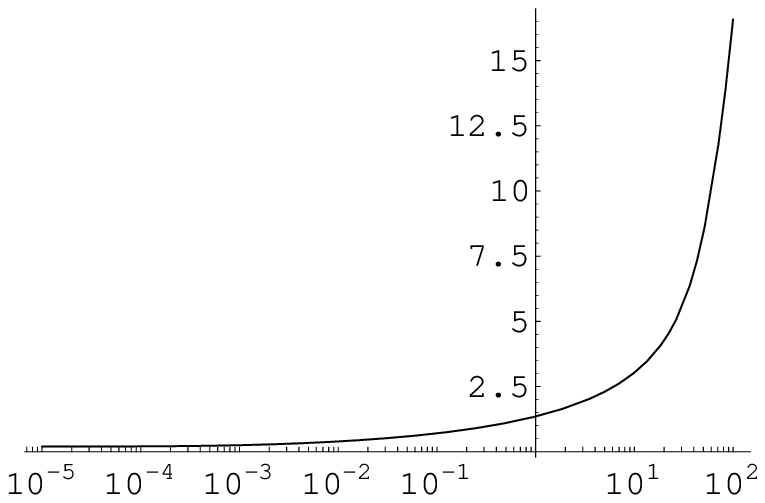,width=10.4cm}} 
  \put(84,70){$y_{\rm res}$}
  \put(128,2){$|\dmeV|$}
\end{picture}
  \caption{The resonance momentum $y_{\rm res}$ at the freeze-out
temperature of the $n$-$p$ interactions, $T=1\MeV$, as a function of
$|\dmeV|$. The initial asymmetry $\eta_i$ and momentum $y_i$ only have
significant effects for very low $|\dmeV|$; we have neglected them in this
plot.}
\label{fnal_2}
\end{figure}

If we want to investigate the influence of the neutrino asymmetry on
BBN, we need to check its value at the temperature $T=1 \MeV$ when the
neutron-proton reactions decouple and the $n/p$ ratio freezes
out. 
In Fig.~\ref{fnal_1} we present the dependence of the maximum possible
asymmetry at the temperature 1 MeV from $|\dmeV|$. The solid line
shows the maximum asymmetry for the initial conditions $y_i=0$ and
$\eta_i=0$. Our result is in good agreement with previous extensive
calculations \cite{ASF99b}. Naturally, we do not expect any dependence
on $\sin2\theta$, since we have assumed full MSW transition. We should
therefore stress that our result is only a good approximation in
regions of the parameter space ($\dm$,$\sin2\theta$) where there is an
effective exponential increase of the asymmetry, and if there is full
MSW transition afterwards.

From Fig.~\ref{fnal_1} we can easily extract that for $|\dmeV| \ll 1$ the asymmetry
will remain small, while for $|\dmeV| \gg 1$ it can reach its
asymptotic value $3/8$, which was found in \cite{fv96} (the factor 2
difference with our $6/8$ on Fig.~\ref{fnal_1} stems from the relation
$\eta = 2 \eta_{\nu_e}$).  For $|\dmeV| < 1$, the asymmetry at
temperature 1 MeV can be fitted according to the simple relation
\be
\eta = 0.1\times |\dmeV|^{3/4}
\ee
in good agreement with the result of \cite{ASF99b}, $\eta = 0.1\times
|\dmeV|^{2/3}$.

We will shortly discuss minor effects on the result. The short dashed
line in Fig.~\ref{fnal_1} shows the effect of a non-zero initial
asymmetry (for $\eta_i=10^{-4}$ in this particular case). A non-zero
initial asymmetry will increase the asymmetry at $T=1\MeV$ for small
$|\dm|$. A non-zero momentum $y_i$ will work in the opposite direction
and decrease the final asymmetry for small $|\dmeV|$ (long dashed line
is for $y_i=0.2$). Note that for very small $|\dmeV|\ll 10^{-5}$ the
exponential regime is to be expected at $T\sim 1 \MeV$ so that our
results become invalid. Finally, the chemical potential plays an
important role when the asymmetry is large. However, its effect is to
exponentially damp the number density of the $\bar\nu_{\rm e}$ (in the
case $\eta>0$), thus decreasing the asymmetry. Therefore, our result
remains the maximum possible asymmetry. We have estimated the decrease
of the asymmetry due to the chemical potential to be less than a
factor of 2.

For illustration, Fig.~\ref{fnal_2} shows the dependence of the dimensionless
resonance momentum $y_{\rm res} = p_{\rm res}/T$ at the temperature $T
= 1$ MeV from the parameter $|\dmeV|$. We  see that at 
$|\dmeV| \ll 1$ the resonance momentum is small, while for 
$|\dmeV| > 10$ it is already on the tail of the distribution
function. This explains the values of the asymmetry presented in
Fig.~\ref{fnal_1}.

\section{Conclusions}

In this work we have found the maximum lepton asymmetry which can be
created at a given temperature due to active-sterile neutrino
oscillations in the Early Universe.  We have assumed full transitions
of active to sterile neutrinos due to MSW resonance to derive a very
simple relation between the lepton asymmetry and the temperature. From
this relation, we found that the evolution of the lepton asymmetry
does not evolve according to a specific power law.

We have also investigated on the maximum asymmetry at the temperature 1
MeV when the $n/p$ ratio freezes out.  We found that the value of the
asymmetry at this time depends on $\dm$, but not on $\sin 2 \theta$ in
concordance with \cite{ASF99b}. Furthermore, the initial settings
described by the initial resonance momentum $y_i$ and asymmetry
$\eta_i$ have only an influence for very small $\dm$. This means that
details of the previous evolution are not very important for the value
of the asymmetry at $T = 1\MeV$. In the parameter region which is
related to a large final asymmetry, $\dmeV > 10$, our asymmetry values
are in good agreement with those of \cite{fv96}.

We stress again that our results depend on the assumption of full
transitions of active to sterile neutrinos. If in some part of the
parameter space the transition is not total, then the asymmetry can be
somewhat smaller. Also, for large asymmetries, the exact value of
$\eta$ may be smaller due to effects from the chemical potential.

\section*{Acknowledgment}
We are grateful to G.~Raffelt for helpful comments and reading the
manuscript. We also would like to thank S.~Hansen, R.~Foot and
R.R.~Volkas for useful comments.

This work was partly supported by the Deut\-sche
For\-schungs\-ge\-mein\-schaft under grant No.\ SFB 375 and in part by
INTAS grant 1A-1065.

%
%
\nc{\advp}[3]{{\it  Adv.\ in\ Phys.\ }{{\bf #1} {(#2)} {#3}}}
\nc{\annp}[3]{{\it  Ann.\ Phys.\ (N.Y.)\ }{{\bf #1} {(#2)} {#3}}}
\nc{\apl}[3] {{\it  Appl. Phys. Lett. }{{\bf #1} {(#2)} {#3}}}
\nc{\apj}[3] {{\it  Ap.\ J.\ }{{\bf #1} {(#2)} {#3}}}
\nc{\apjl}[3]{{\it  Ap.\ J.\ Lett.\ }{{\bf #1} {(#2)} {#3}}}
\nc{\app}[3] {{\it  Astropart.\ Phys.\ }{{\bf #1} {(#2)} {#3}}}
\nc{\cmp}[3] {{\it  Comm.\ Math.\ Phys.\ }{{ \bf #1} {(#2)} {#3}}}
\nc{\cqg}[3] {{\it  Class.\ Quant.\ Grav.\ }{{\bf #1} {(#2)} {#3}}}
\nc{\epl}[3] {{\it  Europhys.\ Lett.\ }{{\bf #1} {(#2)} {#3}}}
\nc{\ijmp}[3]{{\it  Int.\ J.\ Mod.\ Phys.\ }{{\bf #1} {(#2)} {#3}}}
\nc{\ijtp}[3]{{\it  Int.\ J.\ Theor.\ Phys.\ }{{\bf #1} {(#2)} {#3}}}
\nc{\jmp}[3] {{\it  J.\ Math.\ Phys.\ }{{ \bf #1} {(#2)} {#3}}}
\nc{\jpa}[3] {{\it  J.\ Phys.\ A\ }{{\bf #1} {(#2)} {#3}}}
\nc{\jpc}[3] {{\it  J.\ Phys.\ C\ }{{\bf #1} {(#2)} {#3}}}
\nc{\jap}[3] {{\it  J.\ Appl.\ Phys.\ }{{\bf #1} {(#2)} {#3}}}
\nc{\jpsj}[3]{{\it  J.\ Phys.\ Soc.\ Japan\ }{{\bf #1} {(#2)} {#3}}}
\nc{\lmp}[3] {{\it  Lett.\ Math.\ Phys.\ }{{\bf #1} {(#2)} {#3}}}
\nc{\mpl}[3] {{\it  Mod.\ Phys.\ Lett.\ }{{\bf #1} {(#2)} {#3}}}
\nc{\ncim}[3]{{\it  Nuov.\ Cim.\ }{{\bf #1} {(#2)} {#3}}}
\nc{\np}[3]  {{\it  Nucl.\ Phys.\ }{{\bf #1} {(#2)} {#3}}}
\nc{\pr}[3]  {{\it  Phys.\ Rev.\ }{{\bf #1} {(#2)} {#3}}}
\nc{\pra}[3] {{\it  Phys.\ Rev.\ A\ }{{\bf #1} {(#2)} {#3}}}
\nc{\prb}[3] {{\it  Phys.\ Rev.\ B\ }{{{\bf #1} {(#2)} {#3}}}}
\nc{\prc}[3] {{\it  Phys.\ Rev.\ C\ }{{\bf #1} {(#2)} {#3}}}
\nc{\prd}[3] {{\it  Phys.\ Rev.\ D\ }{{\bf #1} {(#2)} {#3}}}
\nc{\prl}[3] {{\it  Phys.\ Rev.\ Lett.\ }{{\bf #1} {(#2)} {#3}}}
\nc{\pl}[3]  {{\it  Phys.\ Lett.\ }{{\bf #1} {(#2)} {#3}}}
\nc{\prep}[3]{{\it  Phys.\ Rep.\ }{{\bf #1} {(#2)} {#3}}}
\nc{\prsl}[3]{{\it  Proc.\ R.\ Soc.\ London\ }{{\bf #1} {(#2)} {#3}}}
\nc{\ptp}[3] {{\it  Prog.\ Theor.\ Phys.\ }{{\bf #1} {(#2)} {#3}}}
\nc{\ptps}[3]{{\it  Prog\ Theor.\ Phys.\ suppl.\ }{{\bf #1} {(#2)} {#3}}}
\nc{\physa}[3]{{\it Physica\ A\ }{{\bf #1} {(#2)} {#3}}}
\nc{\physb}[3]{{\it Physica\ B\ }{{\bf #1} {(#2)} {#3}}}
\nc{\phys}[3]{{\it  Physica\ }{{\bf #1} {(#2)} {#3}}}
\nc{\rmp}[3] {{\it  Rev.\ Mod.\ Phys.\ }{{\bf #1} {(#2)} {#3}}}
\nc{\rpp}[3] {{\it  Rep.\ Prog.\ Phys.\ }{{\bf #1} {(#2)} {#3}}}
\nc{\sjnp}[3]{{\it  Sov.\ J.\ Nucl.\ Phys.\ }{{\bf #1} {(#2)} {#3}}}
\nc{\sjp}[3] {{\it  Sov.\ J.\ Phys.\ }{{\bf #1} {(#2)} {#3}}}
\nc{\spjetp}[3]{{\it Sov.\ Phys.\ JETP\ }{{\bf #1} {(#2)} {#3}}}
\nc{\yf}[3]  {{\it  Yad.\ Fiz.\ }{{\bf #1} {(#2)} {#3}}}
\nc{\zetp}[3]{{\it  Zh.\ Eksp.\ Teor.\ Fiz.\ }{{\bf #1} {(#2)} {#3}}}
\nc{\zp}[3]  {{\it  Z.\ Phys.\ }{{\bf #1} {(#2)} {#3}}}
\nc{\ibid}[3]{{\sl  ibid.\ }{{\bf #1} {(#2)} {#3}}}

\end{document}